\documentclass[preprint,review,12pt]{elsarticle}

\usepackage{etoolbox}
\makeatletter
\patchcmd{\ps@pprintTitle}
  {\let\@oddhead\@empty}
  {\def\@oddhead{\mbox{}\hfill\thepage}}
  {}{}
\makeatother

\usepackage{graphicx}
\usepackage{amssymb}

\biboptions{comma,square,sort&compress}

\journal{~}

\begin{document}

\pagestyle{myheadings}

\begin{frontmatter}


\title{Beach-level 24-hour forecasts of Florida red tide-induced respiratory irritation}


\author[label1]{Shane D. Ross\corref{cor1}}
\author[Clarkson]{Jeremie Fish}
\author[labelK]{Klaus Moeltner}
\author[Clarkson]{Erik M. Bollt}
\author[label2]{~~~~~~~~~~~~Landon Bilyeu}
\author[Mote,NOAA]{Tracy Fanara}
\cortext[cor1]{Email: {\tt shaneross@vt.edu}}
\address[label1]{Aerospace and Ocean Engineering, Virginia Tech, Blacksburg, Virginia}
\address[Clarkson]{Electrical and Computer Engineering and C$^3$S$^2$ the Clarkson Center for Complex Systems Science, Clarkson University, Clarkson, New York}
\address[labelK]{Agricultural and Applied Economics, Virginia Tech, Blacksburg, Virginia}
\address[label2]{School of Plant and Environmental Sciences, Virginia Tech, Blacksburg, Virginia}
\address[Mote]{Mote Marine Laboratory and Aquarium, Sarasota, Florida}
\address[NOAA]{National Oceanic and Atmospheric Administration, Washington, DC}

\begin{abstract}
An accurate forecast of the red tide respiratory irritation level would
improve the lives of many people living in areas affected by algal blooms.
Using a decades-long database of daily beach conditions, two conceptually different models to forecast the respiratory irritation risk level one day ahead of time are trained. 
One model is wind-based, using the current days' respiratory level and the predicted wind direction of the following day.
The other model is a probabilistic self-exciting Hawkes process model. 
Both models are trained on beaches in Florida during 2011-2017 and applied to the red tide bloom during 2018-2019.
  For beaches where there is enough historical data to develop a model, 
the model which performs best depends on the beach.
The wind-based model is the most accurate at  half the beaches, correctly predicting the respiratory risk level on average about 84\% of the time. 
The Hawkes model is the most accurate (81\% accuracy) at nearly all of the remaining beaches.
\end{abstract}


\end{frontmatter}



\section{Introduction}
\label{S:intro}

Harmful algal blooms of the toxic dinoflagellate {\it Karenia brevis},  referred to as 
``Florida red tide'' (henceforth abbreviated as RT) 
have affected the Florida Gulf coast for centuries. 
There is emerging evidence that these blooms have increased in frequency, intensity, and geographic spread in recent years (e.g., \cite{alcock2007assessment,nierenberg2009beaches,nierenberg2010florida,nierenberg2010changes,fleming2011review,corcoran2013primer}). 
{\it K. brevis} produces brevetoxin, a  neurotoxin that can result in massive fish kills and mortalities to marine mammals and sea birds. 
Indirectly, this can lead to neurotoxic shellfish poisoning in humans from consuming contaminated shellfish \cite{kirkpatrick2006environmental,corcoran2013primer}.

More directly, and of primary interest for this study, 
brevetoxin is released into near-shore aerosol as RT cells are lysed by wave action, or aerosolized though bubble-mediated \cite{pietsch2018wind} transport. 
If inhaled by humans, brevetoxin can produce upper and lower respiratory irritation, such as a burning sensation of eyes and nose, and a dry, choking cough. 
While these symptoms have been found to be relatively short-lived in healthy individuals 
(upon separation from the harmful aerosol), 
RT effects can be more severe and longer-lasting for people with chronic respiratory conditions, such as asthma \cite{backer2003recreational,alcock2007assessment,kirkpatrick2009aerosolized,nierenberg2010changes,fleming2011review,kirkpatrick2011aerosolized,corcoran2013primer}.

In this study,  the potential of forecasting beach-specific respiratory irritation one day ahead of time is assessed, using previous irritation reports at the same location. 
Two models are proposed: one based on the current respiratory irritation level and a forecast of the next day's wind direction, and the other based on RT as a self-exciting process (or Hawkes process).
Both are data-driven, trained on data on from  2011-2017, and tested on data from a severe RT bloom during 2018-2019.
Both are compared with a simple persistence model, which assumes the next day's respiratory irritation level will be the same as the current day's. 

\subsection{Monitoring of red tide blooms}

Monitoring for {\it K.\ brevis} blooms involves the following primary components. 
Satellite imagery is  processed to locate potential blooms.
Processed images are made available to managers and state health 
officials in the Gulf of Mexico through the National Oceanic and Atmospheric Administration (NOAA) Harmful Algal Bloom Operational Forecast System (HAB-OFS) \cite{ofshabwebsite}. 
In Florida, where blooms tend to occur most frequently, water samples are collected weekly along the shore and from offshore transects 
by the Florida Fish and Wildlife Research Institute once a bloom is identified. 
Samples are delivered to a laboratory for cell enumeration via microscopy. 
Microscopic enumeration takes about one hour per sample
  \cite{hardison2019habscope}.   
Typically, samples are processed within 1–2 days and can take longer for more samples from remote areas. 
The resulting cell counts are then used by HAB-OFS to provide broad, county-wide forecasts of brevetoxin exposure risks. 
The 
  cell    count data for a particular county can be up to a week old by
the time it is available to the public. 
In terms of forecast accuracy, Stumpf et al.\ \cite{stumpf2009skill} found that 
while county-wide forecasts of respiratory risk were correct 70\% of the time, 
they were only correct  20\% of the time when applied to individual beaches.

Recently, beach-level 24-hour forecasts for respiratory impacts have received the most focus from policy agencies \cite{moeltner2021harmful}.
The Gulf of Mexico Coastal Ocean Observing System (GCOOS) 
recently developed a beach-level risk forecast that   includes more than 20 Gulf Coast beaches \cite{gcooswebsite}. 
The forecast uses current wind forecasts as well as near real-time cell counts of {\it K. brevis} from water samples enabled by HABscope, a portable microscope system \cite{hardison2019habscope}.
While the beach reporting system described below provides different data (actual beach-level respiratory impact, collected daily over several years), 
one can envision the potential to tie in to existing forecast frameworks, fusing the multiple data sources for more accurate forecasts or greater spatial coverage.

\subsection{Beach Conditions Reporting System}

To address the need for location-specific conditions, a Beach Conditions Reporting System (BCRS) was initiated in 2006 
\cite{kirkpatrick2008florida,stumpf2009skill,currier2009ocean,nierenberg2011frontiers}
and was redeveloped in 2015 when it began gaining public usership \cite{bcrswebsite}. 
From 2017-2019, the site gained approximately 1.5 million users. 
The BCRS provided smartphones to (professional) lifeguards and park rangers with an app designed for reporting beach conditions. 
Twice each day (10:00 and 15:00 local time), lifeguards and park rangers report occurrence of coughing as described further below (and other conditions such as presence of dead fish). 
While lacking the quantification and precision of microscopy, the reports provide beachgoers with useful real-time information for adequate planning 
(i.e., severity of aerosolized toxins, potential risks to asthmatics, presence of dead fish, etc.). 
The BCRS 
is managed by Mote Marine Laboratory with lifeguards and park rangers as the primary reporters in several counties \cite{bcrswebsite}. 
The BCRS data compilation 
is automated, with timely sharing of data with agencies, including
Florida Fish and Wildlife Conservation Commission and NOAA.

Though the BCRS provides more timely information about beach conditions than the weekly sampling described earlier, it does not provide key information needed for consistent forecasts. 
It provides no   direct    information on {\it K. brevis} cell presence, but   rather an indirect assessment of {\it K. brevis} cell presence via water color and the presence of dead fish.   

While RT outbreaks can occur throughout the entire year, aerosolized RT impacts have shown substantial variability both in a temporal and spatial sense.
They can last from a few hours to multiple days or even weeks at a given site (e.g., beach), and vary in intensity across sites at a given point in time, with heavily impacted areas at times alternating with completely unaffected shoreline segments 
\cite{nierenberg2009beaches}.

While efforts are ongoing to curb RT blooms via prevention and control methods \cite{alcock2007assessment,vargo2008nutrient,nierenberg2010florida,kirkpatrick2014human}, the predominant management strategy to date has been mitigation, via early detection and avoidance of human contact 
\cite{alcock2007assessment,corcoran2013primer}.

\section{Methods}

Below, the RT respiratory irritation data set is described. Afterward, two different models to predict RT irritation risk one day ahead of time are described: a wind-based model and a Hawkes process model.

\subsection{Respiratory irritation data}

Starting in August 2006, the professional lifeguard corps in Sarasota County began twice-daily reports (approx.\ 10:00 and 15:00 local time) of the presence of respiratory irritation at six sites, as part of the BCRS. 
In January 2007, two additional lifeguard sites were added in Manatee County. 
Respiratory irritation is defined by the amount of coughing observed in addition to the personal conditions experienced by the lifeguard. 
The presence of people coughing is used as a proxy for respiratory irritation (cough, nasal congestion, throat irritation, chest tightness, wheezing, and shortness of breath).
Coughing has been documented as a response to {\it K. brevis} aerosols in studies involving occupationally exposed workers, recreationally exposed beachgoers, and asthmatics 
\cite{backer2003recreational,backer2005occupational,fleming2005,fleming2007}. 

Lifeguards 
`listen' to 
beachgoers for the presence and/or frequency of coughing. 
The symptoms observed by the lifeguards are reported at various levels of respiratory irritation as shown in Table \ref{irritation levels}. 
\begin{table}[!h]
\centering
\begin{tabular}{l l l}
\hline
{\bf Irritation level} &  {\bf Risk level} &
{\bf In a 30 s audio sample}\\
\hline
{\small
None   } & {\small Low} & {\small No coughing/sneezing heard in 30 s }\\
{\small Slight } & {\small Low } & {\small A few coughs/sneezes heard in 30 s }\\
{\small Moderate} & {\small High} & {\small A cough/sneeze heard every 5 s }\\
{\small High  }  & {\small High} & {\small Coughing/sneezing  almost continuously}\\
\hline
\end{tabular}
\caption{The four-tiered red tide respiratory irritation levels reported in the Beach Conditions Reporting System, and the corresponding two-tiered risk level defined for this study.}
\label{irritation levels}
\end{table}
Irritation levels are given on a four-tiered scale from `None' (no coughing noted nearby), `Slight' (a few coughs and sneezes within 30 seconds), `Moderate' (A cough/sneeze heard every 5 seconds), and `High' (continuous coughing and sneezing in nearby surroundings). 
For portions of the analysis below,
moderate and high classes are together as `high' risk, as these were the level at which impacts affect the general public \cite{stumpf2009skill}. 
The `none' and `slight' are grouped as `low' risk.
The two-tiered (binary) classification of RT respiratory irritation risk was used in this study for an initial analysis of forecasting methods, based on historical data.

\subsection{Statistics describing the data set}

\begin{figure}[h!]
\centering
\includegraphics[height=0.55\textwidth]{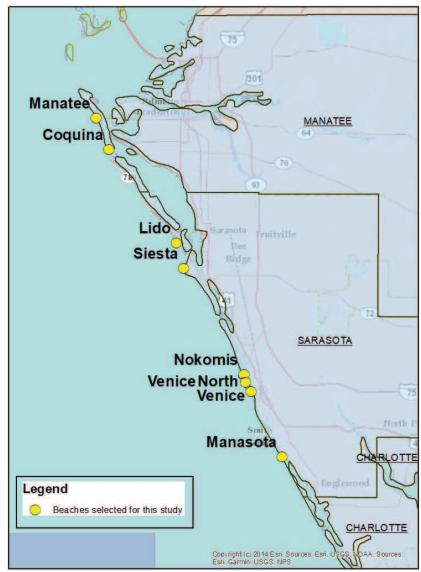} 
\caption{\footnotesize 
Eight beaches along the Florida Gulf Coast out of 40 for which reporting is available from 2011-2019. These eight have the highest number of reported red tide respiratory events: six are in Sarasota County and two are in Manatee County.
}
\label{bcrs-overview}
\end{figure}

\begin{figure}[!t]
\centering
\includegraphics[height=0.35\textwidth]{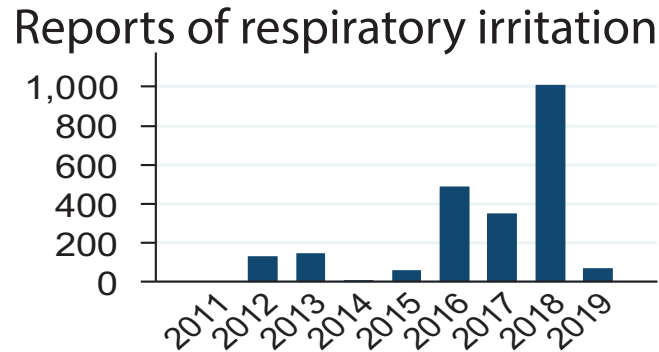}
\caption{\footnotesize 
Reports of respiratory irritation caused by red tide for Sarasota County, from December 2011 to April 2019.
}
\label{bcrs-overview-reports}
\end{figure}

Since December 2011, the BCRS has monitored RT conditions at over 40 beaches in nine counties along the Gulf coast via citizen scientists, 
in most cases local lifeguards.
This study considered only the time frame when 40 beaches were reporting. 
The eight beaches which are the focus of this study (see Figure \ref{bcrs-overview}) were chosen because they were 
the only beaches with enough reports of high respiratory risk days to develop a convergent model, as described in the Methods.
Figure \ref{bcrs-overview-reports}
shows the annual aggregate sum of RT irritation reports (level `slight' or higher) from the six beaches in Sarasota County.
It is  clear that in 2018, the county was especially hard-hit by RT. 
For the purposes of the model, only the highest daily irritation level is considered. Reports are given twice per day, separated by only a few hours,  and are often the same. 
When they differ,  the higher of the four-tiered irritation level is used
to describe the risk-level for that day. 
For the two-tiered irritation risk of high and low, there was near unanimous agreement between the two daily reports.

\begin{figure}[h!]
\centering
\begin{tabular}{cc}
\includegraphics[height=0.4\textwidth]{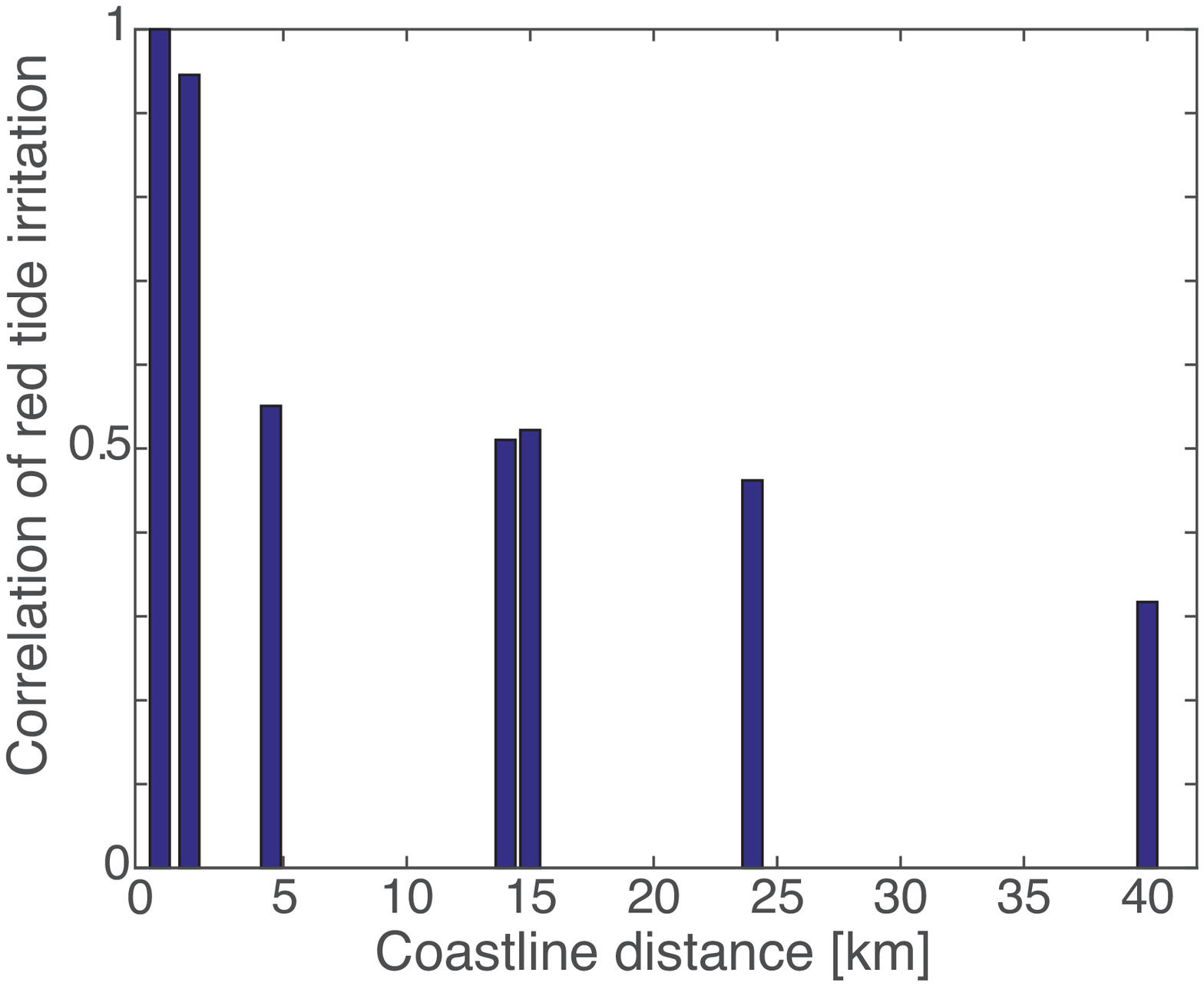} &
\includegraphics[height=0.4\textwidth]{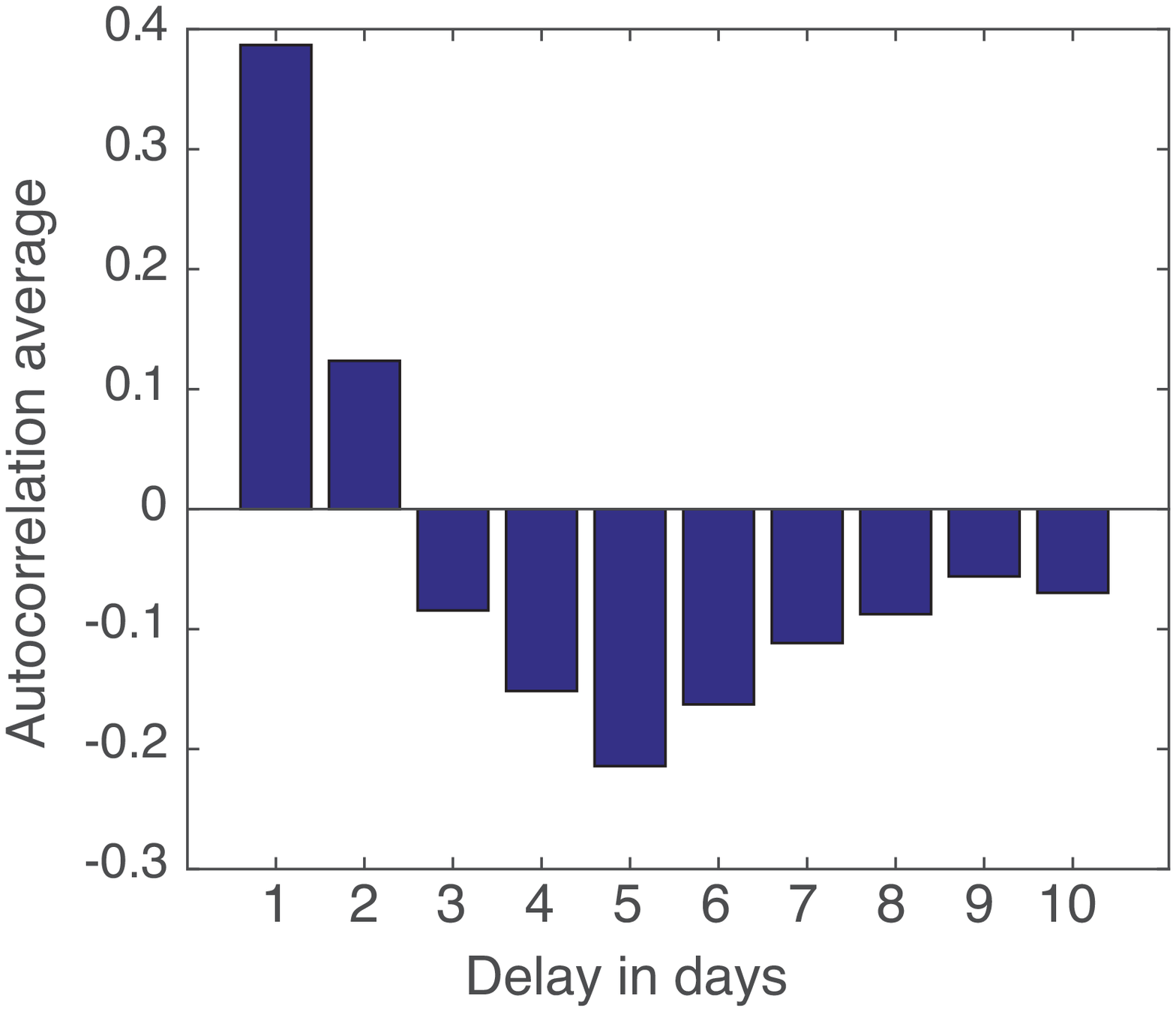}
\end{tabular}
\caption{\footnotesize 
(left) Correlation of red tide respiratory irritation level between the different beaches in Figure \ref{bcrs-overview} as a function of coastline distance.
(right) 
The average auto-correlation of a beach's red tide respiratory irritation level with time.
}
\label{correlation}
\end{figure}

\paragraph{Correlation of irritation level across distance and time}
From the BCRS database,  correlation of the respiratory irritation level with nearby beaches is considered. 
For the four-tier irritation levels described as None to High (see Table \ref{irritation levels}), numerical values 1 through 4 were assigned, respectively. 
As one can see in the left panel of Figure \ref{correlation}, the correlation of nearby beaches (within a few km) is high, 
but drops off to about 50\% for beaches from 5 km up to approximately 25 km away, decreasing approximately monotonically with distance.
One can also consider the correlation of irritation level across time for an individual beach.
The average auto-correlation of a beach's respiratory irritation level versus time-lag is reported. 
Notice there is a moderate correlation for 1 day, but the correlation drops to about zero after 2 days. 
These spatiotemporal correlations are reported merely to describe the data set. They are not incorporated into the models described below.

\subsection{Partition of data into training and testing sets}

The BCRS database contains respiratory irritation reports from December 2011 to April 2019.
Following standard practice in machine learning \cite{brunton2019data}, a data set must be
partitioned into  a training and testing set. A model is constructed from the training set 
and validated on the testing set.
Training on earlier data in order to test on later data is a common method as it mimics forecasting, and this approach will be followed in the current study.
The data reveal that there was a time-frame of a significant number of reports during 2018 until early 2019 (Figure \ref{bcrs-overview-reports}). The 2018-2019 time-frame therefore emerges as a testing data set.
All earlier data, from 2011-Dec-31 to 2017-Dec-31, are considered as the training data set.

The database has long stretches of no respiratory irritation reports (that is, recorded as None).
For forecasting purposes, the ability to predict respiratory irritation during a cluster of such events is a primary focus of this study.
To be precise, a {\it red tide  respiratory irritation episode} is defined on a beach-by-beach basis for all data as follows. A RT respiratory irritation episode is said to begin the day when the BCRS irritation report (see Table \ref{irritation levels}) first goes above None and ends 7 days after the last irritation report above None. 


\subsection{Wind-based model}

\begin{figure}[h!]
\centering\includegraphics[width=0.5\linewidth]{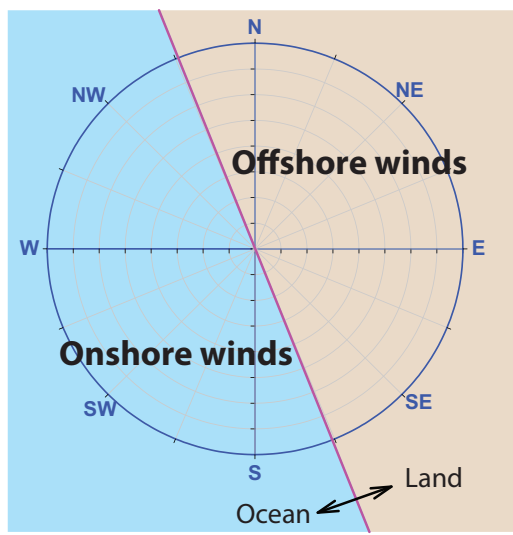}
\vspace{-4mm}\caption{\label{winds_diagram} Definition of onshore and offshore wind directions for the beaches studied. The standard wind direction convention is used, where direction denotes where the wind is coming from.}
\end{figure}

The wind direction reported from the BCRS at each beach was given as one of the   eight usual directional  octants of width 45$^{\circ}$ (N, NE, E, SE, S, SW, W, and NW). 
For the beaches of interest on the west coast of Florida, onshore winds were defined as winds blowing from 168.75$^{\circ}$ to 326.25$^{\circ}$ (S clockwise to NW, following \cite{stumpf2003monitoring,stumpf2009skill}), which presumes the 330$^{\circ}$ to 150$^{\circ}$ orientation of the coastline, as shown in Figure \ref{winds_diagram}. 
It is noted that the beach coastlines do not deviate enough from this assumed coastline for the coarseness of the input wind direction to make a difference in modeling.

\begin{figure}[h]
\centering\includegraphics[width=1\linewidth]{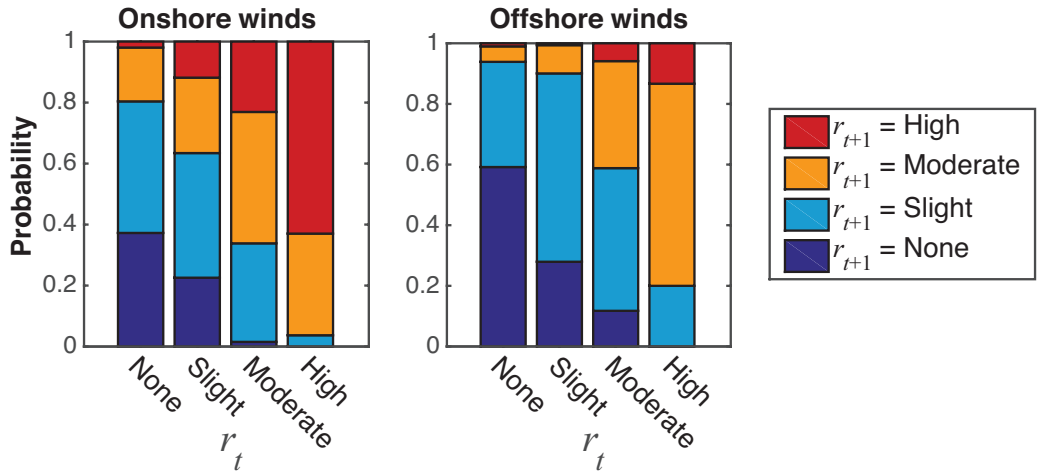}
\vspace{-15mm}\caption{\label{respiratory_irritation_probabilities} Respiratory irritation level on day $t+1$ as a function of the level on day $t$ and the wind condition on day $t+1$; onshore or offshore winds.}
\end{figure}

For a wind-based model, the following statistical analysis from the BCRS database is performed for the training set (years 2011-2017), restricted only to RT respiratory irritation periods. 
For a respiratory irritation level of $r_t$ on day $t$, where $r_t\in\{$None, Slight, Moderate, High$\}$,  the probability of $r_{t+1}$ on day $t+1$ is calculated, based on the frequencies of such occurrences in the database. 
Two conditions are considered: onshore winds on day $t+1$ and offshore winds on day $t+1$. 
The results are shown in Figure
\ref{respiratory_irritation_probabilities}.
For instance, if today's respiratory irritation level is Moderate, and there will be onshore winds tomorrow, then with over 60\% probability, tomorrow's irritation level will be Moderate or High (high risk). On the other hand, if tomorrow's winds are offshore, then with about 60\% probability the irritation level will improve, to Slight or None (low risk).

\begin{figure}[!b]
\vspace{-3mm}
\centering\includegraphics[width=0.7\linewidth]{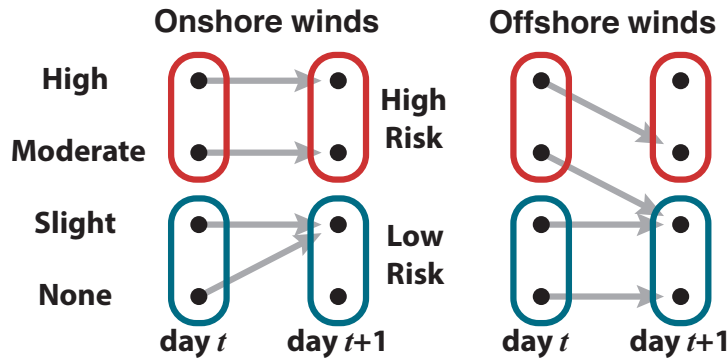}
\caption{\label{wind_model} Wind-based respiratory irritation model.}
\end{figure}

If  this probability distribution is assumed to hold for future events, then one has a straightforward forecast model for the respiratory irritation level based on the weather forecast, in particular, the  wind direction forecast. 
To get a deterministic model in place of a probabilistic model, one can assume that the state with the maximum likelihood is the one which occurs. 
This leads to the simple model given in Figure \ref{wind_model}.
For instance, if the irritation level today is at None, but tomorrow has onshore winds, the maximum likelihood is that it will be Slight tomorrow,
since the light-blue bar in the first stack of the left panel of Figure \ref{respiratory_irritation_probabilities} is longer than all others.
In place of the four-tiered irritation level, one can use the simpler two-tiered irritation risk level, low and high, as given in Table \ref{irritation levels}. 
This coarser, binary description of the irritation level is used as it is more amenable to modeling and validation. 

\subsection{Hawkes process model}
\label{sec:Hawkes_model}

Hawkes processes have been used to model a remarkable range of phenomena. From earthquakes \cite{ogata1988,turkyilmaz2013}, to gang related violence \cite{mohler2011,park2019} to econometrics \cite{errais2010,bacry2013}. This type of model performs well in situations where there is evidence for a clustering of events in time. 

Hawkes processes are a type of temporal point process. The classic temporal point process is the \textit{homogeneous} Poisson process, an example of which is the number of asteroids striking the Earth. Presumably the probability of an asteroid strike is independent of whether or not there was a previous strike. Poisson processes are thus ``memoryless'' and events are roughly equally spaced in time.
An illustrative example of a homogeneous Poisson process compared to a non-homogeneous (Hawkes) Poisson process is provided in 
\ref{app:Example}.

A homogeneous Poisson process is a stochastic point process in which events happen with {\it constant} rate $\lambda$
and with probability,
\begin{equation}
    P_{\rm Poisson}(k|\lambda) = \frac{(\lambda t)^k}{k!}e^{-\lambda t}, \label{eq:PoissonProb}
\end{equation}
where $\lambda$ has units of inverse time. 
A \textit{non-homogeneous} Poisson process is one in which the rate parameter $\lambda(t)$ is a function of time. 
A linear Hawkes process with an exponential kernel \cite{hawkes1971} is a non-homogeneous Poisson process where,
\begin{equation}
    \lambda(t) = \lambda_{0} + \sum \limits_{i} \alpha e^{-\beta(t-\tau_i)}. \label{eq:HawkesRate}
\end{equation}
where $\lambda_0 \geq 0$ is the background intensity rate, $\alpha > 0$ is the excitation level, $\beta > 0 $ is the reversion level, and $\{\tau_1, \tau_2, ... \tau_i \}$ is the observed true sequence of past event times (high risk days in the current study).

A Hawkes process is known as a {\it self-exciting process} because when an excitation happens the rate increases before decaying to the natural unexcited rate $\lambda_0$ (see \ref{app:Example}). The rate initially increases by the amount $\alpha$ when an event arrives but exponentially decays with rate $\beta$ towards $\lambda_0$. 
The Hawkes process maintains a finite rate $\lambda(t)$ so long as $\alpha<\beta$ \cite{laub2015}. 

Estimation of the parameters for the Hawkes process is done via maximum likelihood estimation (MLE) 
of Fisher \cite{fisher1922mathematical}, a procedure for estimating model parameters from a training set such that  the observed data is most probable \cite{brunton2019data}.
Details of MLE for a Hawkes process can be found in \ref{sec:Hawkes}. 
For each beach, the parameters $\lambda_0$, $\alpha$, $\beta$, and $P_{\rm threshold}$ are estimated based on the training data set for that beach. 
If the parameters converge to the same values independent of random initial parameters in the MLE process,  
the Hawkes model has {\it converged}.
Only beaches which have a convergent Hawkes model are included in the Results. 
It is expected that not all, or even a majority of the beaches will have convergent Hawkes models, as
it is known that for some types of data, parameter estimation is quite sensitive to the random initial guess \cite{veen2008}.



Once the parameters $\{\lambda_0,\alpha,\beta,P_{\rm threshold}\}$ have been estimated for a given beach,
the probability of observing no events (high risk days) at each time is determined by 
substituting ${\lambda}(t)$ into eq.\ (\ref{eq:PoissonProb}), where ${\lambda}(t)$ is given by eq.\ (\ref{eq:HawkesRate}).
This gives, on each day $t$, the probability of the next day, $\Delta t = 1~{\rm d}$, being a high risk day, which is $1-P_{\rm Poisson}(0|{\lambda}(t))$, that is,
\begin{equation}
    P_{\rm highrisk} (t) = 
    1 - e^{-{\lambda}(t)\Delta t}.
    \label{eq:p_highrisk}
\end{equation}
The model predicts that the next day, $t+1$, will be a high risk day if
\begin{equation}
    P_{\rm highrisk} (t) > P_{\rm threshold}.
    \label{eq:p_threshold}
\end{equation}


\section{Results}
\label{S:results}

The Hawkes process model parameter estimation was performed for the training data (years 2011-2017) at all 40 beaches in the BCRS database.
The parameter estimation process converged at only 8 of the 40 beaches. They
were located in a geographic cluster (see Figure \ref{bcrs-overview}) and from north to south are
Manatee, Coquina, Lido, Siesta, Nokomis, Venice North, Venice, and Manasota.
The set of parameters are calculated independently for each beach, and reported in
Table \ref{tab:MLEBeaches}. 
\begin{table}[!h]
\footnotesize
    \centering
    \begin{tabular}{|l|c|c|c|c|}
    \hline
       {\bf Beach}  & $\lambda_0$ & $\alpha$ & $\beta$ & $P_{\rm threshold}$ \\ 
       \hline
        Manatee 
       & 0.0027 & 0.2393 & 0.3151 & 0.2932 \\
       \hline      
       Coquina 
       & 0.0039 & 0.3181 & 0.4016 & 0.3451 \\
       \hline
       Lido 
       & 0.0054 & 0.1324 & 0.2332 & 0.2600 \\
       \hline
       Siesta 
       & 0.0059 & 0.2370 & 0.3700  & 0.3680 \\
       \hline
       Nokomis 
       & 0.0041 & 0.1039 & 0.1218  & 0.3137 \\
       \hline
        Venice North    
        & 0.0058 & 0.1457 & 0.1917  & 0.3211 \\   
       \hline
       Venice 
       & 0.0052 & 0.1449 & 0.2404  & 0.2785 \\
       \hline
       Manasota 
       & 0.0049 & 0.2219 & 0.3182  & 0.3350 \\
        \hline \hline
{\bf Average} &  ${\bf 0.0047}$ & ${\bf 0.1929}$ & ${\bf 0.2740}$  & 0.3143 \\
\hline
(Standard Deviation) & (0.0011) &       (0.072)~~  &  (0.094)~~   & (0.036) \\
\hline
    \end{tabular}
    \caption{Estimated values of the Hawkes parameters for each of the beaches considered, listed north to south. Units of  $\lambda_0$, $\alpha$, and $\beta$    are inverse days, ${\rm d}^{-1}$.
   The bottom rows shows the average parameter value (across all beaches) along with 
   the standard deviation.
    }
    \label{tab:MLEBeaches}
\end{table}
In all the beaches examined, the parameter $\beta \approx 0.27 ~{\rm d}^{-1}$, which suggests a correlation time-scale of $\beta^{-1} \approx 3.6$ days. 
A longer probabilistic ``memory'' timescale can be estimated,
\begin{equation}
T_m = -\log(0.05)/\beta + 1 \quad {\rm (in~days)},
\end{equation}
which is about 12 days, 
since a high risk day which 
happened 12 days ago has 
a contribution of about 5\% compared to that of high risk day which occurred 1 day ago. 
Also of interest is that the estimates show that $\alpha \approx 0.2 ~{\rm d}^{-1}$ for all beaches. 
This is roughly $10^3$ times the base rate, 
$\lambda_0 \approx$ 0.005, for each beach, meaning that 
a single high risk day  after several low risk days    increases the probability of more high risk days by several orders of magnitude. 



The wind-based and Hawkes process models were applied to the test data at each of the eight beaches for which a convergent Hawkes process model was found.
For comparison, a persistence model is included as a null hypothesis. 
The persistence model assumes that tomorrow will be like today.
The results are reported in Table \ref{tab:model_results}. 
\begin{table}[!h]
\vspace{-4mm}
\begin{tabular*}{\textwidth}{ l | l  c  c  c }
\hline
{\bf  \scriptsize Beach and Time-frame} & {  \scriptsize Model} &  {  \scriptsize Accuracy} & {  \scriptsize False Negative Rate} & {  \scriptsize False Positive Rate}  \\
\hline 
{\scriptsize Manatee, 181 days } \vspace{-4mm}\\ {\scriptsize 2018-Aug-03 to 2019-Jan-30}
\vspace{-4mm}
&  {\scriptsize Persistence }  & {\scriptsize 77\% } & {\scriptsize 44\% } & {\scriptsize 16\% } \\ 
\vspace{-4mm}
& {\scriptsize Wind        }  & {\scriptsize 78\%   }& {\scriptsize 69\%  }&  {\scriptsize 5\% } \\ 
& {\scriptsize Hawkes     }   & {\scriptsize 80\%  } & {\scriptsize 42\% } & {\scriptsize 13\% } \\
\hline
{\scriptsize Coquina, 181 days  } \vspace{-4mm}\\ {\scriptsize 2018-Aug-03 to 2019-Jan-30} 
\vspace{-4mm}
&{\scriptsize  Persistence } & {\scriptsize 75\%  } & {\scriptsize 46\% } & {\scriptsize 18\% } \\ 
\vspace{-4mm}
& {\scriptsize Wind    }      & {\scriptsize 75\%  } & {\scriptsize 68\% } &  {\scriptsize 9\% } \\ 
& {\scriptsize Hawkes   }     & {\scriptsize 78\%  } & {\scriptsize 44\% } & {\scriptsize 14\% } \\
\hline
{\scriptsize Lido, 239 days   }  \vspace{-4mm} \\ {\scriptsize 2018-Jun-06 to 2019-Jan-30} 
\vspace{-4mm}
& {\scriptsize Persistence }  & {\scriptsize 85\% }  & {\scriptsize 36\% } & {\scriptsize 10\% } \\ 
\vspace{-4mm}
& {\scriptsize Wind       }   & {\scriptsize 83\%   }& {\scriptsize 62\%  }&  {\scriptsize 5\%  }\\ 
& {\scriptsize Hawkes  }      & {\scriptsize 83\%  } & {\scriptsize 56\%  }&  {\scriptsize 6\%  }\\
\hline
{\scriptsize Siesta, 237 days  } \vspace{-4mm}   \\ {\scriptsize 2018-Jun-08 to 2019-Jan-30} 
\vspace{-4mm}
& {\scriptsize Persistence }  & {\scriptsize 83\% }  & {\scriptsize 36\% } & {\scriptsize 12\% } \\ 
\vspace{-4mm}
& {\scriptsize Wind      }    &{\scriptsize  84\%  } & {\scriptsize 43\%  }&  {\scriptsize 8\%  }\\ 
& {\scriptsize Hawkes    }    & {\scriptsize 85\%  } & {\scriptsize 46\% } &  {\scriptsize 6\% } \\
\hline
{\scriptsize Nokomis, 239 days    } \vspace{-4mm} \\ {\scriptsize 2018-Jun-06 to 2019-Jan-30} 
\vspace{-4mm}
& {\scriptsize Persistence }  & {\scriptsize 80\%  } & {\scriptsize 44\% } & {\scriptsize 13\% } \\ 
\vspace{-4mm}
&{\scriptsize  Wind        }  & {\scriptsize 82\%  } & {\scriptsize 60\%  }&  {\scriptsize 5\%  }\\ 
& {\scriptsize Hawkes    }    & {\scriptsize 74\%  } & {\scriptsize 65\%  }& {\scriptsize 14\%  }\\
\hline
{\scriptsize Venice North, 239 days  } \vspace{-4mm}   \\ {\scriptsize 2018-Jun-06 to 2019-Jan-30} 
\vspace{-4mm}
& {\scriptsize Persistence  } &{\scriptsize  80\% }  &{\scriptsize  43\% } &{\scriptsize  13\% } \\ 
\vspace{-4mm}
& {\scriptsize Wind        }  & {\scriptsize 81\%  } & {\scriptsize 61\%  }& {\scriptsize  6\% } \\ 
& {\scriptsize Hawkes    }    &{\scriptsize  75\%  } & {\scriptsize 64\%  }& {\scriptsize 13\% } \\
\hline
{\scriptsize Venice, 238 days     } \vspace{-4mm}\\ {\scriptsize 2018-Jun-05 to 2019-Jan-22} 
\vspace{-4mm}
& {\scriptsize Persistence  } & {\scriptsize 79\% }  &{\scriptsize  47\% } &{\scriptsize  13\% } \\ 
\vspace{-4mm}
& {\scriptsize Wind          }& {\scriptsize 85\% }  &{\scriptsize  53\%  }&  {\scriptsize 4\%  }\\ 
& {\scriptsize Hawkes     }   & {\scriptsize 78\%  } & {\scriptsize 59\% } & {\scriptsize 11\%  }\\
\hline
{\scriptsize Manasota, 232 days    } \vspace{-4mm} \\ {\scriptsize 2018-Jun-07 to 2019-Jan-30} 
\vspace{-4mm}
&{\scriptsize  Persistence  } & {\scriptsize 81\% }  & {\scriptsize 43\%  }& {\scriptsize 13\% } \\ 
\vspace{-4mm}
& {\scriptsize Wind         } &{\scriptsize  87\%  } & {\scriptsize 48\%  }&  {\scriptsize 3\% } \\ 
& {\scriptsize Hawkes     }   & {\scriptsize 78\%  } & {\scriptsize 57\%  }& {\scriptsize 11\% } \\
\hline
\end{tabular*}
\caption{\footnotesize For each beach,  the percent of days for which the respiratory irritation risk was correctly forecast  by each model during the dates shown is reported as the accuracy. Also given are the false negative rate (also called the miss rate) and the false positive rate. 
}
\label{tab:model_results}
\end{table}
For each beach and each model, the accuracy is reported, or
the percent of days for which the respiratory irritation risk was correctly forecast.  Also given are the false negative rate (also called the miss rate) and the false positive rate.

Considering only the accuracy, the model which performs the best depends on the beach.
For the two northern-most beaches and Siesta, the Hawkes model is the most accurate. For the four southern-most beaches, the wind model performs best.
Only at Lido does the simple persistence model perform best.
But notice that even when a model is the most accurate, it might not have the lowest false negative rate, or percent of true high risk days that were not predicted. At all but the two northern-most beaches, persistence  has the lowest false negative rate.
At all but one beach (Siesta), the wind model has the lowest false positive rate (predicting a high risk day which is in fact low risk).

\begin{figure}[h!]
\centering\includegraphics[width=1\linewidth]{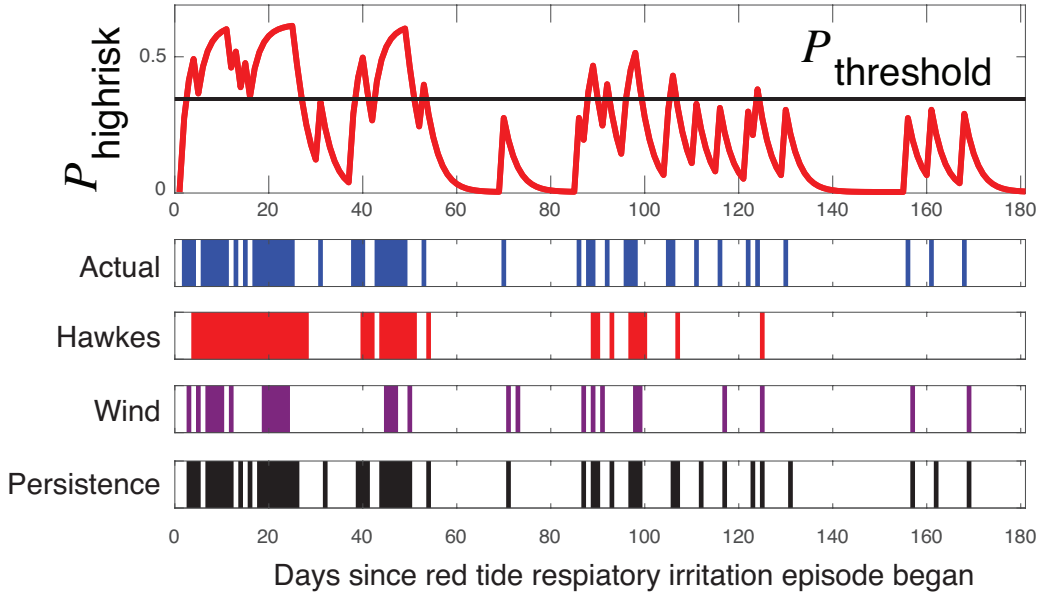}
\caption{\label{coquina} For Coquina Beach, Manatee County, Florida,  the days of actual high respiratory risk level are shown, and several 1-day forecast models, during the red tide respiratory irritation episode that began August 2018. 
The probability of a high risk day using the Hawkes process model is shown at top. The flat black line represents the threshold. 
When the probability of a high risk day exceeds the threshold, the day is forecast as a high risk day. Otherwise, it will forecast as a low risk day. 
The Hawkes model performs best in this case.
The wind-based and persistence models are shown for comparison.}
\end{figure}
\begin{figure}[!t]
\centering\includegraphics[width=1\linewidth]{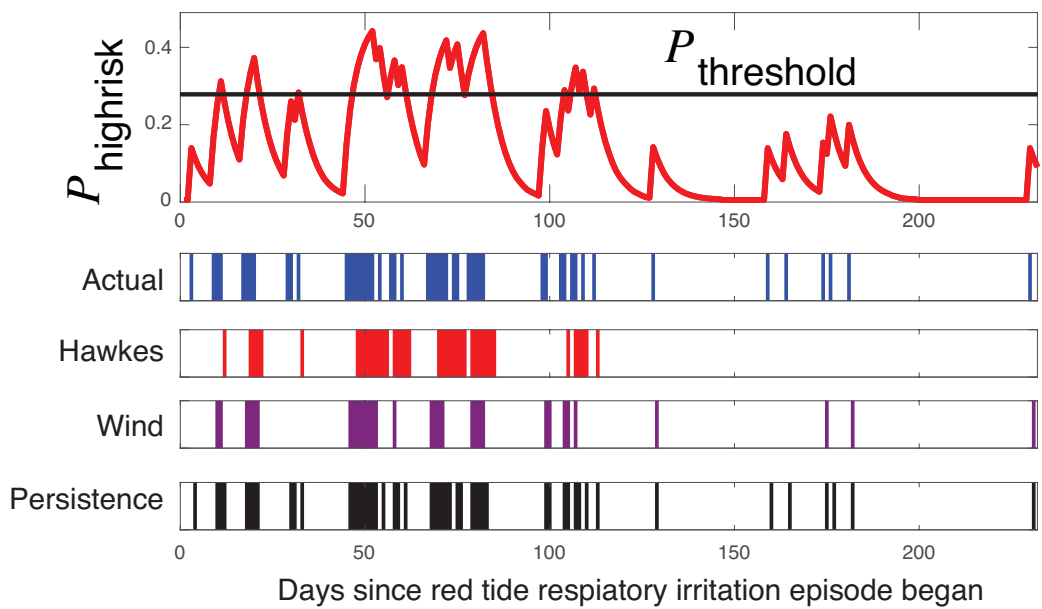}
\caption{\label{venice} Same data as in Figure \ref{coquina} but for Venice Beach, Sarasota County, Florida during the red tide respiratory irritation episode that began June 2018. The wind model performs best in this case.
}
\end{figure}

The self-excitation behavior characteristic of the Hawkes process can be seen in the example of  Coquina Beach,
Figure \ref{coquina}.  
The probability of the following day being high risk is a function of all actual high risk days in the recent past. 
When the probability exceeds the threshold, the model predicts the following day will be high risk day.
For this beach, the Hawkes model had the highest accuracy.
The wind and persistence models are shown for comparison.
Results for Venice Beach are also shown, an example where the wind model had the highest accuracy, in Figure
\ref{venice}.
  
\section{Discussion}

The BCRS has accumulated a significant amount of data since its launch in 2006. 
In the absence of a more sophisticated first-principles-based forecasting system, the historical
data could be utilized to provide a near-term beach-level forecast at the locations considered.
For forecasting the beach-specific  respiratory irritation risk level one day ahead of time,
all the models perform similarly during the 2018-2019 RT respiratory irritation episode, between 75\% and 85\% accuracy (see Table \ref{tab:model_results}).
The simple persistence model (tomorrow will be like today) does rather well, near 80\% accuracy, and at one beach (Lido), was the best performer.
The wind-based model is the most accurate for half the beaches for which a model was developed. All were geographically located near each other, in the southern part of Sarasota County.

Interestingly,
the probabilistic self-exciting Hawkes process model, which does not contain wind as an input, outperforms the more intuitive wind-based model at  nearly half of the beaches.
Fortuitously, there were enough events to get numerical estimates for the parameters of a univariate Hawkes process, but not enough for a multivariate analysis (i.e., including wind data). 
With additional data containing high-risk events, adding wind data to the Hawkes process model may be possible. 
That is, a hybrid of the  models considered here could potentially improve accuracy, but the amount of historical data available to train the model would need to increase significantly.

\paragraph{Limitations}
The given models do not predict when a RT bloom will occur, 
 or when a RT respiratory irritation episode will occur at any given beach.
However, when a 
 RT respiratory irritation episode
   does begin, the models provide a method for forecasting the respiratory irritation risk  for the next day, based on the recent risk history.

Data for all 40 beaches goes back only to December 2011. The total number of high risk days during 
the 
2,193-day training period ranges from a low of zero to a high of 46. Among the eight beaches for which a model was developed, high risk days account for only about 1\% of the training period. This is very little data for which to train a model. With more data, the Hawkes and wind models may improve.

Data from different blooms were lumped together to get enough data to train the models. Yet RT blooms are known to be  patchy and highly variable from bloom to bloom. Thus, the training data may combine behavior from  different blooms with different dynamics. This is a limitation of the data set used.

To simplify the modeling, the four-tiered respiratory irritation scale was coarsened into a two-tiered  risk scale (Table \ref{irritation levels}).
Ideally, respiratory irritation could be measured on a continuous scale, using for example sensors for automatic cough detection and counting at beaches.
Modeling of a continuous variable would increase the number of available data-driven modeling methods.

\section{Conclusions}

The Beach Condition Reporting System which has been operational for over decade, has accumulated a wealth of data. In particular, red tide-induced respiratory irritation levels at individual beaches have been reported daily over this time period. The analysis performed here provides one of the first reports of the statistics of this data set. 

Moreover, beach-level next-day forecasts of the respiratory irritation risk were developed 
on a beach-by-beach basis, if there was enough data in the training set for a model to be developed.
Training  on data from red tide respiratory irritation episodes during the time period 2011-2017, 
 only eight beaches had enough data for a model to be developed (i.e., for parameter values for the probabilistic Hawkes model to converge).
Two types of models provided a forecast of the respiratory level 24-hours ahead of time during the extensive red time bloom of 2018-2019, and were compared with a simple persistence model.
One model was wind-based, using the current days' respiratory level and the predicted wind direction of the following day.
The other model was a probabilistic self-exciting Hawkes process model, which used as input the record of the recent risk history.

No single model performed the best at all the beaches.
The wind-based model performed the best at four of the eight beaches, correctly predicting the respiratory risk level an average of 84\% of the time. 
At three of the eight beaches, the Hawkes model was the most accurate, accurately predicting the next day's risk level an average of 81\%  of  the  time. 
At one beach, the persistence model outperformed both the wind and Hawkes models, with an accuracy of 85\%.
The accuracy of the Hawkes process model at nearly half the beaches for which a model was developed suggests it may be fruitful to consider self-excitation-based approaches in larger-scale models of harmful algal blooms.
Interestingly, the Hawkes process model does not require water samples of {\it K. brevis}, nor ocean or wind forecasts,  and will likely improve by their inclusion.

These results suggest that beach-level on-site reports of respiratory irritation are a valuable data source, providing an excellent means to forecast the following day's beach-specific respiratory irritation risk at the same location. 
Moreover, the efficacy of the BCRS suggests that timely and regular reports of red tide-induced respiratory irritation level should continue to be supported and should be incorporated in  operational forecasts used by resource managers and the public.

\section*{Acknowledgements}
We thank the Mote Marine Laboratory \& Aquarium, the Florida Fish and Wildlife Conservation Commission, the Sarasota County Beach Patrol, the Manatee County Department of Public Safety, the Marine Rescue Division, and all of the beach lifeguards, park rangers, and citizen scientists responsible for providing data into the Beach Conditions Reporting System. 
This project was partially supported by the 
Global Change Center, the Fralin Life Sciences Institute, and the Institute for Society, Culture, and Environment at Virginia Tech.
Moeltner also acknowledges partial support by the USDA/NIFA Multi-State project \#VA-136344. Ross acknowledges partial support by the National Science Foundation (NSF) under grant number 1922516.

\appendix

\section{Example Hawkes Process}
\label{app:Example}

\begin{figure}[h!]
\centering\includegraphics[width=1\linewidth]{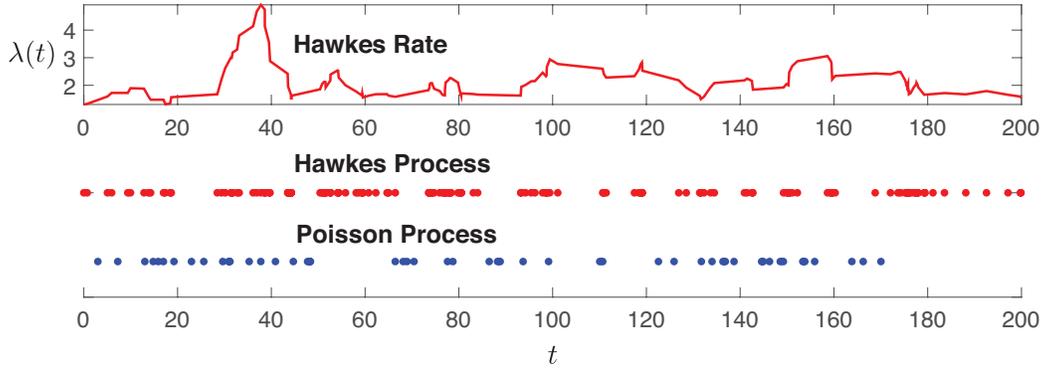}
\caption{The varying rate of an example Hawkes process can be seen (top). A Hawkes process (middle) tends to have events which cluster in time, whereas a Poisson process with the same initial rate (bottom) produces events that are roughly equally spaced in time. The Hawkes process proves to be a good model for many natural processes, which also tend to have events which cluster in time. \label{fig:HawkesVSPoisson}}
\end{figure}

A homogeneous Poisson process (with constant rate $\lambda=1$) is shown in the bottom of Fig. \ref{fig:HawkesVSPoisson}. 
The Hawkes process by contrast is a type of non-homogeneous Poisson process, in which the rate of events are dependent upon the history of arrivals and the time which has passed between events. 
The probability of seeing a new event, e.g., a high risk day, increases when a previous event has occurred. 
This leads to the temporal clustering seen in the middle panel of Fig. \ref{fig:HawkesVSPoisson}. 
A more detailed description of a Hawkes process is provided in Section \ref{sec:Hawkes_model}.

\section{Hawkes Maximum Likelihood Estimation}
\label{sec:Hawkes}
The parameters $\alpha$, $\beta$, $\lambda_0$, and $P_{\rm threshold}$ may be estimated via maximum likelihood estimation (MLE). The log-likelihood function of the Hawkes process is given by \cite{ozaki1979}:
\begin{equation}
    \mathcal{L}(\tau_1,...,\tau_n|\alpha,\beta,\lambda_0) = -\lambda_0 T + \sum \limits_{i = 1}^n \frac{\alpha}{\beta}[e^{-\beta(T-\tau_i)}-1] + \sum \limits_{i = 1}^n \mbox{log}(\lambda_0 + \alpha A(i)), \label{eq:HawkesMLE}
\end{equation}
where $T$ is the total time which has been recorded, 
\begin{equation}
A(i) = \sum \limits_{\tau_i > \tau_j}e^{-\beta(\tau_i-\tau_j)} \ (\forall i\geq 2)
\end{equation}
and $\tau_n$ is the time of the last recorded {\it event}. One may find the partial derivatives of the log likelihood function given in eq.\ (\ref{eq:HawkesMLE}) in \cite{ozaki1979}. However, estimation of the maximum (by setting the partial derivatives equal to zero and solving) is challenging. So as an alternative, standard numerical techniques for nonlinear optimization can be used, in this case the Nelder-Mead direct search technique \cite{nelder1965} to estimate the parameters. 

\end{document}